# Mathematical Sensemaking as Seeking Coherence between Calculations and Concepts: Instruction and Assessments for Introductory Physics

Eric Kuo[1], Michael M. Hull[2], Andrew Elby[3], and Ayush Gupta[3]

[1] University of Pittsburgh, Pittsburgh, PA 15260, USA
[2] University of Vienna, 1010 Vienna, Austria
[3] University of Maryland, College Park, MD 20742, USA

What kind of problem-solving instruction can help students apply what they have learned to solve the new and unfamiliar problems they will encounter in the future? We propose that mathematical sensemaking, the practice of seeking coherence between formal mathematics and conceptual understanding, is a key target of successful physics problem-solving instruction. However, typical assessments tend to measure understanding in more disjoint ways. To capture coherence-seeking practices in student problem solving, we introduce an assessment framework that highlights opportunities to use these problem-solving approaches more flexibly. Three assessment items embodying this calculation-concept crossover framework illustrate how coherence can drive flexible problem-solving approaches that may be more efficient, insightful, and accurate. These three assessment items were used to evaluate the efficacy of an instructional approach focused on developing mathematical-sensemaking skills. In a quasi-experimental study, three parallel lecture sections of first-semester, introductory physics were compared: two mathematical sensemaking sections, with one having an experienced instructor (MS) and one a novice instructor (MS-nov), and a traditionally-taught section acted as a control group (CTRL). On the three crossover assessment items, mathematical sensemaking students used calculation-concept crossover approaches more and generated more correct solutions than CTRL students. Student surveyed epistemological views toward problem-solving coherence at the end of the course predicted their crossover approach use but did not fully account for the differences in crossover approach use between the MS and CTRL groups. These results illustrate new instructional and assessment frameworks for research on mathematical sensemaking and adaptive problem-solving expertise.

## I. PROBLEM SOLVING RELIES ON MATHEMATICAL SENSEMAKING

Research in Mathematics Education has argued that typical classroom problem-solving tasks are not problems at all. Rather, they are *exercises*, designed to help students learn and demonstrate proficiency with specific problem-solving competencies on familiar problem types [1]. As a physics example, ballistic pendulum problems are standard exercises for students to demonstrate competence in applying conservation of angular momentum. On the other hand, true problems present a greater challenge. They are complex, difficult, and time-intensive, and the solutions are not well-practiced ahead of time. The challenge in solving such problems is often in discovering the solution, not just in demonstrating proficiency with established problem-solving skills. Therefore, by definition, there is no standard procedure that guarantees success on all problems. To describe the lack of general problem-solving procedures, Polya said, "To find unfailing rules applicable to all sorts of problems is an old philosophical dream; but this dream will never be more than a dream." [2].

A common instructional approach is to begin instruction by introducing problem-solving procedures for students to follow. These procedures can provide initial scaffolding to help students learn the basic skills they will need. At the same time, we hope that these procedural scaffolds will eventually fade and make way for the creativity, adaptivity, and insight that will help them solve future problems. Significant work remains to be done towards elucidating classroom practices and course materials that can enable such learning, as well as developing assessments that can measure the outcomes of such instruction. Here, assessment poses a fundamental challenge. How do we assess whether students are prepared to solve true problems *in the future* when we can only assess their knowledge and skills *in the present*?

We propose that *mathematical sensemaking* – the practice of seeking coherence between formal mathematics and conceptual understanding – is a key element of effective physics problem solving. Coherence-seeking practices (including, but not limited to, mathematical sensemaking) are central to the problem-solving practices of physicists and engineers [3,4] and well-describe how these professionals develop new insights and find new efficiencies. Additionally, there is evidence that even novice physics learners can engage in such coherence seeking [5,6]. For the long term, we propose that training in mathematical sensemaking can help students look beyond the standard procedures to develop skills useful in tackling the new and difficult problems they will encounter in their academic and professional futures. Local to introductory physics, we argue that mathematical sensemaking is a useful instructional target for physics problem-solving instruction and that operationalizing this instructional target is one of the key issues for problem-solving assessment.

In this paper, we present a new framework for assessing mathematical sensemaking, illustrated through three assessment questions designed to detect some of the novel ways in which introductory physics students can leverage the coherence between formal mathematics and conceptual understanding. Using these assessments to compare two instructional approaches to introductory physics, we will show that students are more likely to learn and engage in mathematical sensemaking if they experience instruction that is intentionally designed to foster such thinking (in conjunction with other PER-based active learning strategies). Our results are novel in two ways. First, the assessments described embody a novel orientation toward operationalizing mathematical sensemaking. Second, the results presented are the first comparison of mathematical sensemaking outcomes for different semester-long instructional approaches in a large-lecture classroom.

## II. MATHEMATICAL SENSEMAKING: SEEKING COHERENCE BETWEEN FORMAL MATHEMATICS AND CONCEPTUAL UNDERSTANDING

In this section, we review the literature on expert and novice problem solving to argue that:

A. Seeking coherence between formal mathematics and conceptual understanding is a normal and explicitly valued aspect of expert physics and engineering practice.
B. Students can and do seek coherence between formal mathematics and conceptual understanding, though it is also challenging for many (if not most) students.
C. Current models for scaffolding and assessing students' problem solving in physics do not adequately capture this coherence-seeking aspect of mathematical sensemaking.

Based on this interpretation of the literature, we propose a new framework for assessing whether and how students are seeking coherence between formal mathematics and conceptual understanding.

### A. Coherence between formal mathematics and conceptual understanding is central to expert physics and engineering practice

Mathematical sensemaking is about building and leveraging coherence between different ideas. The reason to place coherence in such high regard is its centrality to scientific progress and professional science practice. One of the fundamental ways in which the scientific community evaluates new ideas, theories, and experimental results is by considering whether and how they cohere with existing ideas, theories, and experimental results [7,8]. Mature scientific knowledge fits into a coherent structure, and expert knowledge structures mirror this coherence [9]. Conceptual change can be described as the shifting coherence between various ideas, within both scientific disciplines [10] and individuals [11–13].

The practice of mathematical sensemaking has driven discoveries in physics. Dirac's work on relativistic quantum mechanics illustrates how the interplay between conceptual models and mathematics provides a bootstrap for innovation. Dirac generated his relativistic quantum mechanical equation not purely by mathematical manipulation from first principles, but rather by "guessing" a mathematical modification to the Schrodinger equation that preserved the conceptual constraint of Lorentz invariance. This solution was not a formal mathematical derivation, but rather an informal one that relied on connections between mathematics and known physical constraints. Similarly, Dirac's model of a sea of electrons and holes arose by interpreting a surprising mathematical result: negative-energy solutions for the electron. Here, a purely mathematical result led to a new conceptual model, one which formed the basis for the modern understanding of the positron.

As illustrated through Dirac's work, mathematization is often propelled by conceptual innovation. That is, new discoveries can emerge from a scientist's desire (and ability) to generate mathematics that cohere with the physical world, not just from straightforward mathematical procedures. For example, the formulation of Maxwell's equations arose not from a first-principles derivation, but from a novel conceptual analogy treating electric and magnetic fields as rotating gears and free wheels [14]. By applying known mathematical descriptions of mechanical rotations to "electromagnetic rotations," Maxwell developed a mathematical model for electricity and magnetism. At times, the need for coherence between conceptual reasoning and calculations even takes precedence over mathematical formalisms. One example comes from Dirac's use of his delta function, an ill-behaved mathematical entity which was needed to produce physically realizable probability amplitudes [15].

Even outside of extraordinary cases of innovation, the everyday practice of professional scientists and engineers relies on links between formal mathematical calculations and conceptual processes to model novel situations. Gainsburg [3] showed how the interplay between conceptual reasoning and calculation was central to the modeling practice of structural engineers. In trying to understand how forces were being transmitted through a building's structure, the engineers in Gainsburg's study faced a complex problem with no standard analytic procedure. This required them to engage in a modeling cycle, translating different conceptual models of force transfer into an associated mathematical model until they generated a satisfactory description. Similarly, Clement [4] demonstrated that experts justify formal mathematics with informal conceptual strategies, including analogy, limiting cases, and symmetry, in their mathematical problem solving. Clement proposed that experts' mathematical modeling of novel physical situations is grounded in initial conceptual modeling that seeks to describe the phenomenon by aligning known physical principles with informal notions of causality and visual imagery (or what could be thought of as "physical intuition").

### B. Students can productively engage in seeking coherence between formal mathematics and conceptual reasoning

Even as students explore new topics for the first time, they can engage in these same coherence-seeking practices. Even though novel reasoning for a student is rarely a novel addition to society's collective knowledge, these small innovations provide evidence that students can engage in mathematical sensemaking in ways continuous with expert scientific practice. Sherin [16] showed that $3^{rd}$ semester physics students can use their conceptual reasoning to generate novel equations by drawing upon symbolic forms, knowledge elements that tie the general mathematical structure of an equation to a conceptual schema. Rather than a first-principles approach, these students drew upon symbolic forms to create equations representing intuitive conceptual models of physical situations. In one example, two students wrote an expression for the acceleration of a falling ball experiencing air resistance. Their equation, $a(t) = -g + \frac{f(v)}{m}$, where $f(v)$ is the force of air resistance as a function of velocity, expressed their conceptual idea that an "upward acceleration" from air resistance opposed a "downward acceleration" from gravity. Although this formula is consistent with a derivation from Newton's $2^{nd}$ law, students did not appear to use such a derivation, writing this expression directly. Sherin argued that these students generated their equation from the *opposition* symbolic form, which combines the symbol template   -   with the conceptual schema of two influences in opposition. By plugging in mathematical expressions representing the two influences, gravity and air resistance, students were able to express their informal idea of two accelerations in opposition. As knowledge elements blending mathematical structure with conceptual understanding, symbolic forms embody mathematical sensemaking. Additional research on symbolic forms has shown similar instances of students generating novel mathematical expressions to describe physical systems [17], identified efficient problem-solving insights driven by symbolic forms-based reasoning [6], and classified new symbolic forms for other areas of scientific and mathematical reasoning [18].

Students' coherence seeking can also go the other way, using mathematics to sharpen their conceptual understanding. Schwartz, Martin, and Pfaffman [19] found that prompting the use of math to explain the behavior of a balance beam helped young children better recognize the two key physical properties, mass and distance from the pivot point, and even identify torque (mass × distance) as the explanatory physical quantity. Here, the precision of mathematics led students to find more complete conceptual accounts of balance. Along similar lines, Sherin [20] found that two undergraduate physics students spontaneously used a calculation to resolve a conceptual tension between two competing intuitions. These students tackled the question of how the mass of a block traveling at an initial speed $v_0$ would affect the distance it would slide on a rough surface before coming to rest. The students articulated two opposing conceptual effects: a greater mass would (1) decrease the sliding distance by increasing the force of friction and (2) increase the sliding distance by increasing the inertia of the block. To determine the result of these two competing effects, students calculated the acceleration of the block with Newton's 2$^{nd}$ law, yielding $a = \mu g$. In interpreting their calculation and final expression, the students concluded that the two conceptual effects cancel out, making the acceleration of the block independent of the mass. Here, the result of a calculation helped students advance their conceptual knowledge. Along similar lines, Tuminaro and Redish [21] showed that introductory physics students can read out conceptual relationships between physical quantities from equations. In one example, students map mathematics to physical meaning by using Coulomb's law to precisely determine how the amount of charge would have to change in order to keep the electric force constant when the distance doubles. In an experimental demonstration of this type of coherence-driven thinking, Singh [22] found that student performance on qualitative questions could be improved if they were given an isomorphic quantitative problem beforehand. As with Sherin's students, the written responses here indicated that the calculations used on the quantitative problems sharpened students' conceptual reasoning on the qualitative problems.

In sum, students can seek and find coherence between formal mathematics and conceptual reasoning, even as they are learning physics, making this kind of reasoning a viable target for instruction. However, a key difficulty for instructional evaluation is that current assessment paradigms in physics education research are not designed to capture this reasoning, leaving these mathematical sensemaking-related learning outcomes usually undetected and undocumented.

### C. Standard PER assessment paradigms do not fully capture the coherence-seeking aspect of mathematical sensemaking

There are two main types of assessment questions in PER, quantitative questions and qualitative questions, which we argue miss key aspects of coherence-seeking and mathematical sensemaking.

#### 1. *Quantitative problem-solving assessments*

Research on quantitative physics problem-solving instruction has developed step-wise problem-solving procedures for novice students to follow [23–28]. Although the details of the step-by-step sequences differ, the commonalities among the various procedures have established a standard problem-solving paradigm in PER.

1) Describe the physics of the problem – identify the entities and physical processes in the problem.
2) Plan a solution – select the relevant physics principles, express them as equations, and explain how these will be combined with problem-specific features to reach the answer.
3) Execute the solution – Execute the mathematical plan to compute the solution.

4) Evaluate the answer – Check the answer to see if the solution makes sense (e.g. does it have an obviously incorrect sign, magnitude, or units? Are the functional dependences as expected? Do the limiting cases make sense?).

This problem-solving paradigm has successfully addressed one of the primary "coherence problems" tackled by PER: students can solve quantitative problems without engaging with the underlying physical concepts. Instead, their solution approaches can be driven by problem surface features [9] or equations that contain the relevant known and unknown variables [29]. The first two steps of the instructional procedure are designed to focus student attention on the underlying physical entities and principles at play, and instruction emphasizing this initial conceptual analysis helps students engage in conceptual reasoning as a way to select the relevant physics principles and equations [30–32]. Problem solving assessments, when looking beyond just the correctness of the final answer, code for each problem-solving step, capturing where students make errors or deviate from this paradigm in their own problem solving [33]. In sum, this problem-solving paradigm is a major way in which PER has helped students develop coherence-seeking between physics concepts and formal mathematical methods.

While this problem-solving paradigm embodies one important aspect of coherence seeking, mathematical sensemaking does not consist only of conceptual analysis for selection and set up of mathematical equations. Conceptual understanding can also lead to quantitative insight without formal mathematical calculations. Kuo, Hull, Gupta, and Elby [6] showed that students who linked the velocity equation $v = v_0 + at$ to the *base + change* symbolic form, with a conceptual schema that "the ending amount equals the base amount plus a change," used that equation to find a conceptual shortcut on a kinematics problem. This conceptual insight bypassed the need for the formal, mathematical manipulations described by the "execute the solution" step of the standard problem-solving paradigm. Because this form of mathematical sensemaking skips a major step in the problem-solving paradigm, it is not well captured by the associated problem-solving assessments [34]. For this reason, we argue that assessment paradigms that can detect mathematical sensemaking will attend to this critical distinction between calculation and conceptual approaches to reach solutions on quantitative problems.

*2. Qualitative assessment questions*

On the other hand, qualitative questions are used to examine other aspects of students' knowledge described as "functional knowledge" [35] or "conceptual knowledge" (as implied by the fact that qualitative tests are often referred to as "concept inventories"). Investigations with qualitative questions have uncovered students' difficulties with various subjects [36–38]. Students incorrectly reason with the underlying conceptual entities and relations between them, and these errors have been classified as "incorrect intuitions," "pre-conceptions," and "misconceptions."

Comparing the student success rate on qualitative questions to their quantitative problem-solving skill has been an important contrast for PER assessment. Students can have difficulties with qualitative problems that instructors and researchers view as equivalent to, or even simpler than, the quantitative problems they solve in their physics courses [36]. Comparisons of student work on closely related quantitative and qualitative problems show that their success at answering these two types of problems are not equal, with students typically performing worse on the qualitative problems [35,39,40]. Along similar lines, Kim and Pak [41] found little relation between the number of quantitative problems solved in class and performance on the Mechanics Baseline Test, which includes qualitative problems. Researchers commonly interpret these findings as indicating a lack of coherence between students' conceptual and calculation knowledge/skills, with students having weak conceptual understanding and relatively stronger calculation skills.

But while qualitative assessments have revealed a lack of coherence between conceptual understanding and mathematical skill, they again fail to capture dimensions of mathematical sensemaking established in prior research. Students can use formal mathematics to bring precision to their reasoning on qualitative questions, as demonstrated by students in Sherin's [20] and Schwartz, Martin, & Pffafman's [19] studies. Yet, research using qualitative questions is often thought to indicate conceptual understanding only, and typically does not specify whether students use their mathematical calculation skills to answer these qualitative questions. Again, we argue that assessments that can capture these known facets of mathematical sensemaking will distinguish between calculations and conceptual approaches for answering qualitative questions.

To address the challenges raised from quantitative and qualitative problem-solving assessments, we propose a new framework for assessing mathematical sensemaking, one which attends to the use of calculations and conceptual reasoning in students' problem solving.

### D. Calculation-concept crossover: an assessment framework highlighting new aspects of problem-solving coherence

We propose that using calculations on qualitative problems and conceptual approaches on quantitative problems indicates a type of coherence-seeking not addressed by standard problem-solving paradigms in PER. We introduce an assessment framework that looks for *calculation-concept crossover* (Fig. 1), where problem-solving approaches deviate from those typical notions of qualitative and quantitative problem solutions. Certainly, many quantitative questions in physics require a calculation to determine the precise answer and many qualitative questions can be efficiently answered with conceptual reasoning. However, this is not always the case. The calculation-concept crossover framework proposes that the use of crossover approaches, when beneficial, can serve as an indicator of mathematical sensemaking, and hence coherence-seeking practices. This framework thereby builds on the notion of coherence between formal mathematics and conceptual understanding embodied by standard assessment practices.

We argue that calculation-concept crossover indicates both knowledge and epistemological stances supporting coherence. On the one hand, calculation-concept crossover indicates that students have developed knowledge and skills that support fluency with each reasoning approach and the flexibility to apply them for different reasoning tasks. On the other hand, the flexibility illustrated by calculation-concept crossover reveals epistemological stances that support coherence and integration. Prior research on students' epistemologies illustrates the opposite side of the coin, showing how epistemological stances that oppose integration can suppress coherence between different modes of reasoning, even when students are skilled with each mode in isolation [42].

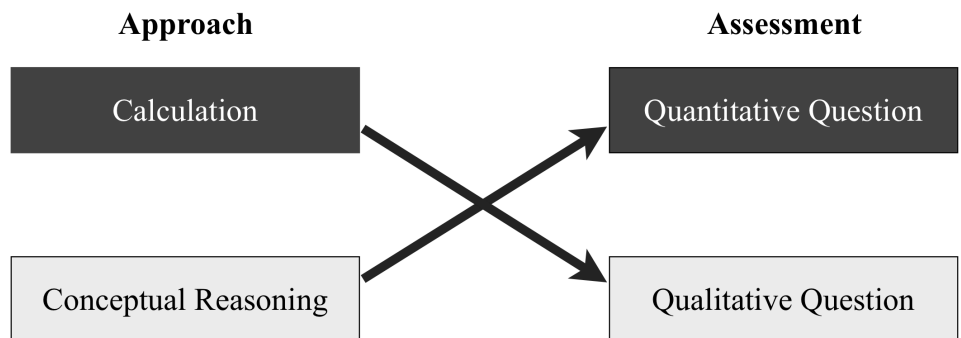

Figure 1. A framework for assessing coherence in how students solve problems, highlighting *calculation-concept crossover*.

We will present three crossover assessment items showing how the increased flexibility provided by introducing calculation-concept crossover in one's "problem-solving toolbox" can lead to more efficient, insightful, and accurate problem solving. These examples will support our central claim, that the crossover assessment framework is generative for creating assessment items that measure a worthwhile aspect of physics problem-solving behavior. Understanding these crossover assessments and the framework critically requires an operationalization of calculation and conceptual approaches.

*1. Operationalizing "calculation approaches"*

*Calculation approaches* are ones that rely on formal mathematical manipulation rules to produce numerical (or symbolic) answers. This is the "execute the solution" step of the problem-solving paradigm, and, colloquially, the "chug" part of "plug-and-chug." Again, as previous PER work in quantitative problem solving has shown, calculations are not devoid of conceptual understanding: conceptual understanding of the physical situation leads to the generation of the appropriate mathematical calculations. However, what we attend to here is whether formal mathematical manipulations are used to produce a final answer. The key operational feature of a calculation for this study is the *eventual reliance on explicit mathematical manipulations to find a solution*, even when conceptual reasoning is attached.

*2. Operationalizing "conceptual approaches"*

On the other hand, *conceptual approaches* here are approaches that avoid formal mathematical manipulations by relying instead on an understanding of *physical concepts* and/or *mathematical concepts*. Use of physical concepts involves reasoning about physical entities and the processes between them. For example, a student could reason that pushing down on a piston adiabatically will increase the energy of the gas inside, because the piston does work on the gas by exerting a force over a distance. Although consistent with a mathematical calculation involving the 1$^{st}$ law of Thermodynamics and the Work-Energy Theorem, here the reasoning avoids explicit mathematical manipulation, and so is labeled as a conceptual approach. Use of mathematical concepts involves reasoning about physical quantities and functional relations between them, as represented in equations. For example, one could reason that if the voltage in a circuit were doubled and the total resistance were halved, then Ohm's law, $I = V/R$, would predict that the current would quadruple. Again, although this answer is consistent with a mathematical calculation and draws on mathematical ideas of proportionality and multiplication, we operationalize this response as a conceptual approach. We distinguish it from an approach where one plugs in the values, explicitly performs the algebraic/arithmetic manipulations, and then arrives at the answer that the current is quadrupled. Instead, in the conceptual approach, the answer is "read out" from the form of the symbolic expression.

In prior research, the careful distinction has been made between physical concepts and mathematical concepts for understanding physical systems [43]. Here we include both as conceptual reasoning, since both contrast with the mathematical manipulations used for calculations. This is the key operational feature of conceptual reasoning for this study: that the conceptual reasoning itself is used to reach the answer *without eventual reliance on explicit mathematical manipulations*.

These operational definitions lead to two key points in classifying solution types. First, the presence or absence of mathematics alone cannot be used to categorize the approach as calculation or conceptual. Physics equations are both representations of the conceptual (i.e., structural/causal) relations between quantities and tools for mathematical manipulations [44]. How an equation is used determines whether it was used for a calculation or conceptual problem-solving approach. Second, incorrect approaches can also be classified as calculations or conceptual reasoning. Incorrect calculations that draw on the wrong equations/values or contain errors in the mathematical manipulations are still calculation approaches. Incorrect conceptual reasoning still embodies a conceptual reasoning approach distinct from calculations.

In the next section, we detail three crossover assessment items to illustrate the potential advantages of crossover approaches. We will then present results of an empirical classroom study that uses these three crossover assessments to measure the problem-solving learning outcomes of two different approaches to teaching introductory physics.

## III. THREE ASSESSMENTS OF CALCULATION-CONCEPT CROSSOVER

### A. Using calculations on a qualitative question: reliability and precision

#### 1. The nature and benefits of this crossover reasoning

Calculation use is an understudied way of improving students' performance on qualitative problems. Consider the typical, qualitative circuit question, shown in Figure 2:

*What happens to the brightness of bulbs A and B when the switch is closed?*

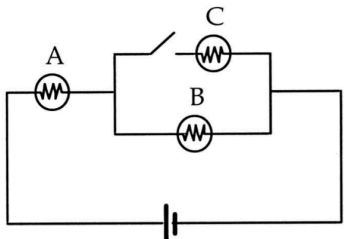

Figure 2. The typical, qualitative circuit question (from Engelhardt and Beichner [45])

Studies of student reasoning on similar DC circuit questions has revealed common errors in conceptual approaches. Students may overgeneralize the rule that parallel branches in a circuit are independent, which would lead them here to incorrectly conclude that the brightness of bulb B will not change [46]. Additionally, students may use local, sequential reasoning to conclude, for instance, that the brightness of bulb A will not change because the current doesn't reach the switch until after it has passed through bulb A. Note, these common errors emerge from reasoning about the physical entities and processes of this circuit, suggesting that many students approach this qualitative question with conceptual reasoning.

Even if students identify the correct conceptual effects of closing the switch, it may be difficult to work out which competing effects have a larger magnitude. When the switch is closed, bulb B receives only a fraction of the total current in the circuit—but the total current increases because the circuit's equivalent resistance decreases. Although the relative magnitudes of these changes are underdetermined by these conceptual arguments, in cases with two opposing effects, students often incorrectly claim that the effects exactly compensate, causing no overall change [36].

Here, calculating the power dissipated in each bulb directly is a more precise approach. Even though the question requires only a qualitative determination of how the brightness of the bulbs change, the mathematical machinery of the calculation determines which of the competing effects wins out. As with Sherin's students, taking a calculation approach, if well-practiced and reliable, can help students avoid conceptual errors and also sharpen their conceptual understanding of the competing effects.

#### 2. Assessment item: qualitative judgment question

In our study, we used the following *qualitative judgment* question to investigate whether introductory physics students will use calculation approaches to answer a qualitative question. The qualitative judgment question is embedded within a set of associated standard questions (Table 1).

Table 1. The qualitative judgment question and associated standard questions

| Assessment Type | Question Text |
|---|---|
| | *Two identical masses, each of mass m = 50 g, are fastened to each other with a bit of plastic explosive. We're going to launch it into the air and detonate the explosive at the highest point. (Ignore air resistance throughout this problem.)* |
| Associated Standard Questions | *(a) Suppose we launch the pair of masses at an angle θ = 60 degrees above the horizontal, from a spring gun. The spring has a spring constant of k = 1000 N/m, and we compress it x = 10 cm (sin 60° = 0.87; cos 60° = 0.5). Find the maximum height of the pair of masses, taking its initial height to be 0.*<br>*(b) At exactly that instant, when it's at the highest point, we detonate the explosive. And it so happens that the instant after the explosion, one mass (A) is not moving at all. Find the velocity of the other mass (B)*<br>*(c) Find the distance between the masses A and B when they hit the ground.*<br>*(e) Sketch a graph of the vertical and horizontal components of the velocity for mass B from the time of launch to the time it hits the ground. Explain your reasoning. You don't need to make precise calculations, just show the shape of the graph in your sketch.* |
| Qualitative Judgment Question | *(d) During the explosion, mass B speeds up while mass A comes momentarily to rest. Does the overall mechanical energy of the two-mass system increase, decrease, or stay the same during that explosion? Explain.* |

     For this qualitative judgment question (part d), the correct answer is that the overall mechanical energy of the system increases during the explosion. Although much of the problem invites calculations, we hypothesize that, even though they know how to calculate the change in energy in this case, students will tend to use conceptual approaches on this qualitative question. On this problem, there are two different conceptual reasoning pathways to the correct answer. In terms of physical entities, one could reason about the energy transfer processes: the chemical potential energy of the explosive is released, doing mechanical work on the two masses and thereby increasing the mechanical energy of the system.[1] Another approach uses the mathematical concepts in the kinetic energy equation: because the mass of the system is effectively halved while the speed doubles, the kinetic energy will increase because the proportional dependence of KE on speed is greater than the dependence on mass (KE ~ $v^2$ vs KE ~ m). However, as with the circuit question described previously, students may also take incorrect conceptual approaches. A student could overgeneralize a commonly stated rule: the law of conservation of energy states that energy is always conserved, so the mechanical energy stays the same. Also, a compensation argument could be used to incorrectly conclude that the total mechanical energy stays the same, because the amount of kinetic energy lost by mass A could be exactly balanced out by the gain in KE of mass B. We predict that some students taking a purely conceptual approach will use some of these common, incorrect arguments.

     Again, a calculation can offer precision and safety from common conceptual errors. Here, because the qualitative judgment question is embedded in a series of quantitative problems, numerical results from previous parts can be used to explicitly calculate the pre-explosion and post-explosion kinetic energies. From part b), the speed of masses A and B is 5 m/s immediately before the explosion and $v_A = 0$ and

---

[1] This reasoning is correct for the class of problems where the explosion increases the speeds of both blocks in the center-of-mass frame without changing the speed of the center of mass in the rest frame, as in this problem. Without this condition, chemical potential energy can be used to decrease mechanical energy, such as when a rocket uses its engine to slow down. Because the assessment item only asks about this case and no other, we did not demand that students explicitly describe this special condition to be considered correct in their reasoning.

$v_B$= 10 m/s immediately after the explosion. Calculating the overall kinetic energy before and after the explosion shows that it increases from 1.25 J to 2.5 J. Because the change in gravitational potential energy is negligible immediately before and after the explosion, the overall mechanical energy increases.

As with the qualitative circuit problem, either conceptual reasoning or a calculation alone is sufficient to reach the correct qualitative answer, that the mechanical energy of the two-mass system increases. However, the assessment target here is coherence-seeking between calculation and conceptual reasoning. Because we do not tell students what approach to take, their spontaneous choices show both the knowledge and disposition for mathematical sensemaking during physics problem solving. Activation of formal calculations on qualitative problems is one of the benefits of mathematical sensemaking that we propose is not well attended to in typical instruction and assessment. Additionally, we believe that calculations will provide an alternative approach that protects students from common, incorrect conceptual arguments, increasing accuracy on this question. Of course, since both approaches must cohere, using both conceptual and calculation approaches together can further increase accuracy by checking that both give the same answer.

### B. Using conceptual reasoning on quantitative problems: finding efficient shortcuts by recognizing conceptual similarity

*1. Nature and benefits of this crossover reasoning*

Wertheimer (1945) asked 6$^{th}$-grade students to solve arithmetic problems of this type: (283+283+283+283+283)/5 = ?. Although students could solve the problem correctly because they had learned addition and long division, many used an explicit mathematical calculation, even though an efficient, conceptual insight can be used here. Kuo, Hull, Gupta, and Elby [6] illustrated this same phenomenon in introductory physics through the following problem:

> *Suppose you are standing with two tennis balls on the balcony of a fourth-floor apartment. You throw one ball down with an initial speed of 2 meters per second; at the same moment, you just let go of the other ball, i.e., just let it fall. What is the difference in the speeds of the two balls after 5 seconds—is it less than, more than, or equal to 2 meters per second? (use g = 10 m/s$^2$ and neglect air resistance)*

In contrast to calculating the speeds of each ball after 5 seconds to find that the difference in speeds is 2 m/s, some students found an efficient shortcut: the difference in speeds after 5 seconds will be the same as the initial difference, 2 m/s, because both objects gain the same amount of speed over 5 seconds. Even though the formal calculation will yield the correct result, what is notable about this conceptual approach is that it provides an elegant, insightful answer that bypasses the need for a calculation.

Another example of conceptual approaches leading to efficiency and insight comes when solving a series of isomorphic questions. Consider the following pair of problems:

*Linear momentum collision: A block of mass M is initially traveling at a speed $v_0$ when it collides with another block of mass 3M which is initially at rest. After the collision, the two blocks stick together, traveling at the same speed. What is the final speed of the two-block system?*

*Angular momentum collision: A disk of mass M and radius R is initially rotating at an angular speed of $\omega_0$ when it collides with another disk of mass 3M and radius R which is initially at rest. After the collision, the two disks stick together, rotating at the same speed. What is the final angular speed of the two-disk system?*

A common conceptual structure exists for these two problems. Initially, the (angular) momentum is all in the object of mass M. After the collision, the (rotational) inertia/mass of the system increases by 4x, so the (angular) speed must decrease by 4x in order for (angular) momentum to be conserved. By noticing the common conceptual structure of these two problems, the solution of one problem can be directly mapped onto an isomorphic one without additional calculation. As in the case of Wertheimer's arithmetic problem and Kuo et al.'s kinematics problem, a conceptual approach here provides an efficient, elegant way to avoid explicit calculations on a quantitative problem.

*2. Assessment item: isomorphic calculation questions*

In our study, we investigate whether students take conceptual approaches on quantitative problems with *isomorphic calculation questions* about a block on a ramp (Table 2):

Table 2. The isomorphic calculation questions and associated standard question

| Assessment Type | Question Text |
|---|---|
|  | *A block of mass M sits on a ramp of angle θ.* |
| Associated Standard Question | (a) First, suppose the block is frictionless and is held in place by a light string extended parallel to the surface of the ramp, as shown here. Write an expression for the magnitude of the tension in the string. 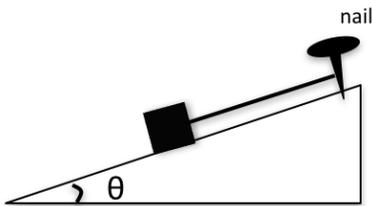 |
| Isomorphic calculation Questions | (b) Instead of a peg, we have the cord connect over a pulley to another block. The second block is just the right mass so that the first block remains at rest. Write an expression for the magnitude of the tension in the string. 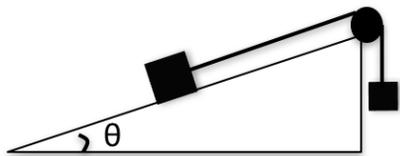 |
|  | (c) Now, suppose there's no string, but the block stays in place because of friction (with coefficient of static friction μ) between the block and the ramp. Write an expression for magnitude of the friction force by the ramp on the block. 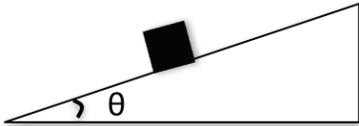 |

For all three questions, the correct calculation uses Newton's 2$^{nd}$ law to calculate the forces along the surface the ramp. The force *F* on the block directed up the ramp is due to the tension in the string or static friction. The force down the ramp is the component of gravity on the block parallel to the ramp, *mg sin θ*. Since the block stays in place, its acceleration is zero. Newton's 2$^{nd}$ law, *ΣF = ma,* yields *F = mg sin θ*.

After the initial question, subsequent questions can be answered by pointing out the conceptual isomorphism between the questions: in all cases, the component of the block's weight down the ramp is balanced by a force pointing up the ramp. Therefore, if the answer for the up-the-ramp force in part a) is *mg sin θ*, then the answer to subsequent parts must also be *mg sin θ*. Instead of performing an identical calculation repeatedly, this conceptual insight uses the isomorphism between problems to efficiently obtain the solution.

How is this shortcut driven by coherence-seeking between calculations and concepts? By identifying the conceptual isomorphism between the three problems, we know that the resulting calculations must also yield identical answers, *even without doing those calculations again*. Finding this shortcut on these ramp problems can evidence a disposition favoring elegant, efficient insights over standard procedures, as well as the skills for correctly identifying such an insight.

### C. Mapping math to meaning: Detecting conceptual errors in a symbolic expression

#### 1. Nature and benefits of this crossover reasoning

A third benefit of coherence between formal mathematics and conceptual reasoning is the ability to relate calculation results to a conceptual interpretation. Physicists often view symbolic answers as superior to numerical answers, because the symbolic answer explicitly represents relationships between the quantities. Similarly, students are able to "[map] mathematics to meaning" in physics problem solving [21]. Additionally, checking the physical meaning of a quantitative result is consistent with the "check your solution" part of the standard physics problem-solving paradigm, although other instantiations of checking the solution are often foregrounded. For example, some checks only specify units, signs, and magnitude; others broadly state: check whether your solution is reasonable and answers the question.

Here, we investigate whether students check the functional dependencies of a symbolic expression with the expected physical behavior. This is a conceptual evaluation of a solution to a quantitative (and, in this case, symbolic) problem. One way to assess this practice is to see if students spontaneously perform these comparisons after obtaining a symbolic answer to a quantitative problem. However, the student's behavior on classroom tasks may depend largely on time constraints, students' expectations about "what the professor wants," and other such factors. Partly for this reason, we decided to assess calculation-concept crossover *skill* instead of *proclivity + skill*. To accomplish this, we engineered a problem that explicitly directs students to crossover approaches, rather than looking for their spontaneous use.

#### 2. Assessment item: cued symbolic evaluation question

Instead of first asking students to calculate a symbolic solution to a quantitative problem, we ask them to evaluate the solution to a quantitative problem without performing the relevant calculation (Table 3). Rather than searching for a spontaneous evaluation of symbolic expressions, this *cued symbolic evaluation question* directly asks for it. By disallowing calculations and explicitly requesting an evaluation of an expression first, this item tests students' skill at using conceptual reasoning to debunk an incorrect, proposed solution. We label this as a crossover approach, because we do not allow students to use a calculation to evaluate a quantitative solution.

Table 3. The cued symbolic evaluation question and associated standard problem

| Assessment Type | Question Text |
|---|---|
| | 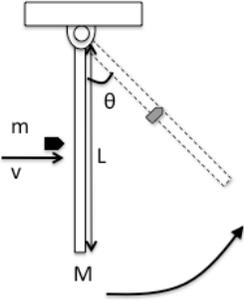 A uniform rod of length L (= 1.00 meter) and mass M (=1.80 kg) is hanging vertically from a frictionless pivot at its top end. A bullet of mass m (=400 g) strikes the rod at the center of the rod and gets embedded in it (See figure 3). Right at the instant before the bullet hits the rod, the velocity of the bullet was entirely horizontal (and perpendicular to the rod) and the magnitude of the bullet's velocity was v (=100 m/s). You can imagine that the rod with the embedded bullet would rotate about the pivot. The moment of inertia of the rod about the pivot is $\frac{1}{3}ML^2$. |
| Cued Symbolic Evaluation Question | (a) Solving for the angular speed of the rod with the embedded bullet immediately after the collision, a student comes up with this answer: $\omega = \frac{3Mv}{2mL}$. Is that a plausible answer? Explain your reasoning to her without solving for that angular speed yourself. |
| Associated Standard Question | (b) Now solve for the angular speed of the rod with the embedded bullet immediately after the collision. |

Here, two valid pathways (in our data set) correctly lead to rejection of the proposed equation for angular speed. Both involve comparing the mathematical expression to expected physical outcomes. One pathway is to reject the mathematical dependencies as not reflecting the physical dependencies of the system. For example, a larger rod mass should resist the motion more, yet the proposed mathematical equation says that angular speed $\omega$ increases as the rod's mass $M$ increases. Similarly, a larger bullet mass $m$ will cause the rod to move faster after the collision, but the proposed equation says that the angular speed decreases as the bullet's mass increases. The other pathway is to compare speeds before and after the collision. Because $v_f = \omega\frac{L}{2} = \frac{3M}{4m}v$ and $\frac{3M}{4m} > 1$, this equation implies that the speed of the bullet increases after the collision, which violates both conservation of momentum and common sense. Here, dimensional analysis alone will not detect the errors in this expression since it has the correct units.

## IV. MATHEMATICAL SENSEMAKING INSTRUCTION TO FOSTER COHERENCE BETWEEN CALCULATIONS AND CONCEPTS

We hypothesized two impedances to crossover approach use. First, developing the knowledge and skills to leverage the coherence between calculations and concepts across qualitative and quantitative problems is difficult. Second, typical instructional approaches may not emphasize this coherence, leading to formation of epistemological views that calculations and conceptual reasoning are distinct. In introductory physics courses, this epistemological messaging could arise in several ways: quantitative and qualitative questions can be seen as separate kinds of problems, quantitative calculations can be perceived to be "real" physics, time constraints caused by content coverage demands restrict the time needed for students to engage in deep coherence-seeking between calculations and conceptual reasoning, and so on. Indeed, as Hammer (1994) documented, students immersed in these instructional environments can view physics as consisting of disconnected pieces and problem solving as requiring formula selection and manipulation, views that likely impede the development of calculation-concept crossover and other flexible problem-solving approaches.

Here, we present a *Mathematical Sensemaking* (MS) instructional approach, developed to help combat the content-based and epistemological challenges to seeking coherence during physics problem solving. We hypothesized that the MS instruction was a better instructional approach for fostering coherence seeking and problem-solving flexibility in introductory physics. To investigate this hypothesis, we conducted a quasi-experimental classroom study using the three calculation-concept crossover assessments described earlier to compare learning outcomes of the mathematical sensemaking instruction to those of a more traditionally-taught course. On the one hand, this was a test of the mathematical sensemaking curriculum, to see if it could produce measurable benefits compared to traditional instructional methods. On the other hand, using these assessments provided a test of whether the calculation-concept crossover assessment framework could be operationalized to provide a useful measurements of mathematical sensemaking in introductory physics students' problem-solving behaviors.

The mathematical sensemaking instruction draws on common pedagogical techniques from educational research. For example, the large lecture incorporates peer-instruction-style clicker questions and peer discussion [39,47]. The instructors use the clicker questions to generate student discussion of ideas. Importantly, for problems where responses do not converge to the correct answer, the instructor will elicit explanations for the two or three most popular choices. The classroom motto is to figure out not only why the right reasoning is right, but also why the wrong explanations are wrong. This aims to give students experience with resolving inconsistencies in the service of developing more coherent understandings. Additionally, the instruction focuses on developing students' epistemologies for coherence-seeking between calculation and conceptual reasoning and contains explicit epistemological messaging along these lines, as described in more detail by Redish and Hammer [48]. Next, we elaborate on two classes of instructional strategies used in this curriculum used to foster mathematical sensemaking.

### A. Explicit strategies to build coherence between mathematics and conceptual understanding

One focus in the mathematical sensemaking was to explicitly help students engage in coherence between mathematics and conceptual understanding. For example, when introducing equations, the instructor would often give a symbolic form interpretation of the equation, explicitly stating the conceptual schema that fits with the mathematical structure. Additionally, problems were designed to elicit both conceptual and calculation approaches, especially on problems for which we predicted these two modes of reasoning would not be consistent. These questions further prompted students to resolve inconsistencies between the two modes of reasoning, to further help students refine their problem-solving strategies and make them aware that they should seek this coherence in physics. As an example, consider this problem from the second homework assignment:

> *Standing on a cliff, I take one rock and throw it straight up at a speed of 30 m/s. I take another rock and throw it straight down at 30 m/s. Suppose the cliff is 50 meters high.* (see Fig. 3)
>
> a) *Just based on common sense, which rock would be moving faster when it hits the ground, 50 meters below? What's the reasoning for that?*

b) Now find an answer based on the kinematics of constant acceleration: Find *x(t)* and *v(t)* for each of the rocks, and find their respective speeds when they hit the ground 50 m below the point of release.

c) Did your answer to (b) agree with your answer to (a)? If not, try to reconcile the contradiction: Figure out what it is about the reasoning in part (a) or part (b) that doesn't work. Get it all to make sense!

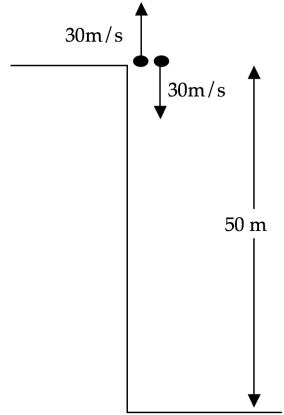

Figure 3. Diagram for "Rocks and Cliff" problem.

Part (a) asks students for their intuitive, conceptual reasoning. Students commonly respond that the rock thrown downward will hit the ground faster. One intuition is that the initial downward motion of the rock adds on to the effects of gravity, whereas the initial motion of the rock thrown upward opposes gravity, so the rock thrown down will hit with a greater speed. In part (b), students used kinematic equations for $x(t)$ and $v(t)$ to calculate the speed of each rock at the bottom of the cliff. The correct calculations show that the rocks land with the same speed. Part (c) asks students to resolve potential disagreements in parts (a) and (b). One conceptual resolution here is that motion under gravity is symmetric: even though a ball tossed upward at 30 m/s is moving away from the final destination, when it returns to its initial height, that ball will be traveling 30 m/s downward. By demonstrating and addressing a common disagreement between conceptual reasoning and calculations, this question aims to help student bring these two forms of reasoning into alignment by developing students' conceptual understanding and their tendency to seek and value such alignment.

The theme of aligning intuitions with calculation serves as a persistent thread in the mathematical sensemaking course. Even by the second homework assignment, students have seen this type of resolution modeled in class and have attempted it in class and on homework. Through this repeated practice, seeking such resolutions becomes the normal problem-solving activity in the class. Although these novice students may have difficulties generating the correct physics resolution in part (c), the homework solutions present a resolution after students have tried to reach one on their own. This three-part problem structure is a good example of the type of reasoning the mathematical sensemaking instruction aims to teach. As the course goes on and explicit scaffolding fades, the hope is that students will routinely seek this type of coherence while learning physics.

### B. Framing physics as independent thinking, not standard procedures: an example from lecture

In this section, we briefly present a snippet of physics instruction that, we argue, supported students' mathematical sensemaking. This example comes from an implementation of the mathematical sensemaking instruction outside of our study, by an instructor who is experienced with the approach. The course was a calculus-based introductory physics course of about 65 students. The excerpt shows how a clicker question in lecture expanded to a larger discussion around students' ideas.

The question presented Xena and Yara, standing 30 meters apart on frictionless ice. Yara has twice the mass of Xena. Xena pulls on a rope that Yara has tied around her waist. On an in-class clicker question and a previous homework question, the students addressed questions about the relative speeds of the two people after Xena tugs the rope, and where they would collide—all assuming a massless rope. They had found that Xena would move twice as fast: $V_{Xena} = 2V_{Yara}$.

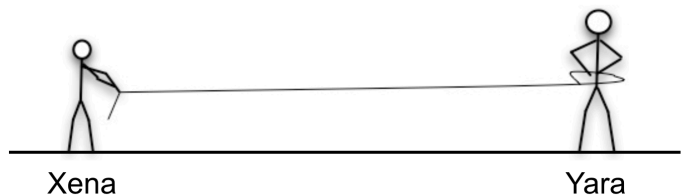

Xena                                          Yara

Figure 4. Xena and Yara standing on frictionless ice

At the start of the excerpt, after the clicker question and subsequent discussion, a student, Brian, asked, "In these types of problems they always say the mass of the rope doesn't matter. Why would the mass of the rope make a difference?" While it is a standard problem-solving assumption in introductory physics that is often glossed over, since the mass of the rope is often much less than the other objects in the problem, the instructor took the opportunity to promote this student's question rather than provide an immediate explanation. The instructor hadn't planned to address this scenario, but he decided to make a new clicker question on-the-fly so that the class could consider Brian's issue:

*How do the post-tug speeds of Xena and Yara compare, assuming a rope of non-negligible mass?*

(A) $V_{Xena} = 2V_{Yara}$
(B) $V_{Xena} > 2V_{Yara}$
(C) $V_{Xena} < 2V_{Yara}$

After students discussed the question in small groups and voted with their clickers, the conversation resumed:

*Instructor:* You've already talked to people, so you've heard arguments. 50% say B and then there are answers in other categories. So, let's hear arguments for B. Why would you say B? Somebody who believes B.
*Oona:* Well, when the rope's mass didn't matter, the force that got to Y is the same as the other [inaudible], but now the weight matters, so some of the force goes into accelerating the rope, so there's less force that gets to Y. So, the acceleration is [inaudible].
*Instructor:* So, some of the force that Xena exerts doesn't get to Y because some of that force goes into accelerating the rope, was that argument for B.

In his revoicing, the instructor ignores Oona's conflation of weight and mass to highlight the mechanism she expresses: because some of Xena's pulling force "goes into accelerating the rope," less force is available to act on Yara. The instructor then opens the class the further arguments, and Michael responds, building on Oona's idea of considering the rope's motion:

*Michael:* Um, I put C because if you look at what happens to the rope with now, non-negligible mass, as Xena and Yara move close to each other, the rope is going to move towards the point where they meet--wherever that may now be. So, whatever force is on Xena

*Instructor:*    and whatever force is on Yara, doesn't just move them. It moves them and some portion of the rope.

*Instructor:* Ok...

*Michael:* And so, I'm assuming that they're still going to end up closer to Yara's starting point than Xena's. But even if they don't, even if they end up right smack dab in the middle, if you add the mass of half of the rope to Xena, it has a greater effect than adding the mass of half the rope to Yara. Because Yara is more massive to begin with. So, the rope is less a percent of her mass. So if you divide the force on them by that new mass, it will result in a greater change in Xena's acceleration, which means that her final velocity will be less than two times Yara's final velocity.

*Instructor:* Awesome. Your argument, I think, if I understand your argument is--let's suppose, let's think of the rope as in two halves. And let's just assign this half of the rope to Xena and this half of the rope to Yara and let's consider these as two objects. Adding half of the mass to Yara doesn't make as much of a difference to Yara's mass as adding half of the rope to Xena makes to Xena's mass. Is your argument. So we should have more of an effect on Xena's mass, therefore more of an effect on her velocity. Wow, all right. This is good. This is good. So, so...yeah, I'm torn over what to do with this right now. But I, so, so...can you respond to an argument that you disagree with?

    In these few minutes of classroom interaction, the instructor makes several instructional choices aimed to support mathematical sensemaking. First, instead of simply answering Brian's question about why assigning the rope a non-negligible mass makes a difference in these sorts of problems, he treated the question as one that could lead to productive conceptual and/or mathematical sensemaking, "promoting" it to the status of clicker question. Next, in revoicing Oona's idea, he focused attention not on the correctness or incorrectness of her account, but on her intuitive story of why the rope's mass could affect Yara's motion. This may have helped create space for Michael to speak up and express his reasoning, which "plays the same game" of considering the rope's motion and how it takes away from Yara's motion—and Xena's motion as well, by Michael's argument. Then, while revoicing Michael's argument, the instructor offers positive comments ("Awesome…Wow, all right. This is good."). This sends the message that Michael's reasoning, which blends cause-and-effect conceptual reasoning about the rope's motion with proportion-based reasoning about why half-a-rope's worth of mass added to Xena has a bigger effect (percentage-wise) than half-a-rope's worth of mass added to Yara, is valued in the class. Note that Michael's reasoning was incorrect: the center-of-mass of Xena's (or Yara's) "piece" of the rope does not accelerate at the same rate as Xena (or Yara), and hence the mass of that piece cannot simply be added to Xena's (or Yara's) mass. But supporting Michael's mathematical sensemaking took precedence in this moment, over promotion only of fully-correct physics.

    This example illustrates a more general pattern in the lecture instruction of the mathematical sense-making oriented lecture instructors in this study. Their instruction included not only pre-planned opportunities for mathematical sensemaking, but also (i) in-the-moment recognition of emergent opportunities for mathematical sensemaking and (ii) the broadcasting of messages that such reasoning is valued. These moves could help create a classroom climate supportive of the epistemological stance that seeking coherence between calculations and concepts is possible and productive.

### C. Predictions of mathematical sensemaking students' performance on crossover assessments

    In this study, we pose our three crossover assessment items to students in a Control (CTRL) lecture section and Mathematical Sensemaking (MS) lecture sections of the same physics course. The Control class emphasized conceptual understanding and quantitative problem solving of the type typically emphasized in end-of-chapter textbook problems, with class discussions in lecture arising from student questions. There was no clear indication of PER-based instructional methods being used in the Control class. Overall, our prediction is that the Mathematical Sensemaking instruction fosters coherence between

calculations and concepts, coherence that does not automatically develop through standard instructional approaches, and therefore the Mathematical Sensemaking students will use more crossover approaches than the Control students do. We also predict that, after instruction, Mathematical Sensemaking students will express epistemological views that more strongly favor coherence between mathematics and concepts in problem solving, i.e., a stronger *MS epistemology*. Importantly, we predict that crossover use and MS epistemology will be positively correlated: students who espouse coherence views will be more likely to demonstrate calculation-concept coherence in answering physics questions.

For each of the three crossover problems, we also investigate whether use of crossover approaches increases the correctness of students' answers:

*Qualitative Judgment Question (Exploding blocks in mid-air)*: In light of the common conceptual reasoning errors possible, we find it reasonable to predict that, compared to CTRL instruction, the MS instruction will increase correctness on this tricky qualitative problem by increasing calculation use on this problem. As with Sherin's students, the calculation may bring increased precision.

*Isomorphic Calculation Questions (Blocks on ramps)*: Because these types of problems can be solved with simple calculations, the predicted increase in conceptual reasoning approaches of the MS students will demonstrate insight and efficiency, but perhaps not increased accuracy.

*Cued Symbolic Evaluation Question (Ballistic pendulum equation evaluation)*: Even though the item *tells* students not to use calculations, we predict that, compared to CTRL students, MS students will use conceptual methods of evaluating mathematical expressions more often, leading to more correct evaluations.

## V. METHOD

### A. Participants

Participants were undergraduate students enrolled in a first-semester calculus-based introductory physics course, taken mostly by engineering majors, at a large, public, research university. Over 15 weeks, the weekly class time consisted of 2.5 hours of lecture in a large lecture hall led by an instructor and a 50-minute discussion section led by a TA. 347 students across three course sections consented to have their data used in this study. Consent rates were relatively low for the CTRL section (56%) as compared to two mathematical sensemaking sections taught by physics education researchers (94%).

### B. Design

Because of large enrollment, students at this university were split between three different sections of the course, each with a separate lecture instructor, discussion sections, homework assignments, and midterm exams. The Control class was taught by a theoretical physicist. One class using the Mathematical Sensemaking (MS) curriculum was taught by a senior physics education researcher. Both of these instructors had taught in this department for at least ten years and were regarded as excellent instructors. Another course using the MS curriculum was taught by a junior physics education researcher, who was teaching a large lecture course for the first time (MS-nov), though they had taught smaller courses using research-informed instruction methods. The two Mathematical Sensemaking classes used the same lecture materials and homework assignments. The instruction in the CTRL course was not affected by this study. The CTRL instructor taught as they normally would. (Note: we are using "they" as the gender-neutral pronoun for all instructors).

The primary comparison of interest is between CTRL and MS groups, demonstrating what can be achieved by two experienced instructors, each using their respective teaching approaches. A comparison between the MS-nov and CTRL groups is of secondary interest, to investigate possibilities for first-time, large-lecture instructors to accomplish the novel goals of the mathematical sensemaking curriculum.

The key assessments occurred at two points. A set of crossover assessments were included on a common final exam, co-designed by the three instructors. Each of the three free-response problems included one crossover assessment item—qualitative judgment, isomorphic calculation, or cued symbolic evaluation—as a sub-part, attached to associated standard problems. The test also contained 10 multiple-choice items, which are not included in our analysis. Students in all three instructional groups took the 2-hour final exam simultaneously. As described below, we separately coded students' responses to the crossover items and to the associated standard problems.

In addition, a modified expectations survey (MPEX2) [49] was given during the last week of class. This survey contained 29 items, most from the MPEX2 and some created to target the Mathematical Sensemaking curriculum's explicit goals. We took 15 items from this survey related to math-concept coherence and seeking coherence during problem solving to construct an *MS epistemology score*. Students completed the survey online, outside of class. There was no systematic pre-survey given. See Supplemental Material at [URL will be inserted by publisher] for the 15 items used to construct the MS epistemology score.

### C. Coding scheme for crossover assessments and the associated standard problems

For the associated standard problems, calculation problems were coded only on whether students used the correct approach. Approaches that correctly plugged problem-specific values into appropriate equations were coded as correct, even if arithmetic errors led to incorrect final answers. Because some sub-parts of each problem were related, answers that correctly utilized incorrect values calculated in previous sub-parts were coded as correct, to avoid multiple penalties for initial errors. For the problem requiring graphs, the graphs were coded as correct if they correctly represented the qualitative behavior.

For each of the three crossover assessments, we coded for (i) the approaches taken and (ii) correctness. The next subsections describe the specific coding scheme for each problem. If a student used multiple approaches, as least one approach appropriately leading to the correct final answer was sufficient to be coded as correct. See Supplemental Material at [URL will be inserted by publisher] for a more detailed discussion of the coding scheme, examples of coded student work, and details on how disagreements between coders were resolved.

#### 1. Qualitative judgment question (exploding blocks in mid-air)

For the qualitative judgment problem, there were approach codes for both *calculation* and *conceptual approaches*. The calculation code is the crossover code on this qualitative question. Solutions that were coded as calculation included plugging numerical or symbolic values into a mathematical expression and then performing mathematical manipulations to compute a final value. The conceptual code indicated justifications for a final answer that did not use an explicit calculation, either by reasoning about physical quantities or reasoning about mathematical dependencies using relevant equations. Since calculation and concept use were independent, students could be coded as attempting both or attempting neither.

Solutions were coded as *correct* if they indicated that the mechanical energy increased and gave a correct justification. The correct calculation involved correctly calculating the kinetic energies before and after the explosion. Correct conceptual reasoning argued that (i) the chemical energy in the explosive was converted to mechanical energy, (ii) the explosion did work on the masses, increasing their kinetic energy, or (iii) halving the mass and doubling the speed would lead to an increase in mechanical energy since kinetic energy depends more strongly on speed than mass (since $KE \sim m$ and $KE \sim v^2$).

*2. Isomorphic calculation questions (block on ramp)*

For the isomorphic calculation problem, we again coded for *calculations* and *conceptual* approaches. Calculation required producing a mathematical expression and performing at least one manipulation or substitution to produce an equivalent expression. Simply writing a mathematical expression or describing a calculation in words was not sufficient to be coded as calculation. A conceptual approach was demonstrated by an explicit statement indicating that the situation was isomorphic to a previous problem, so the solution should be the same as before. As there were two isomorphic calculation questions, the crossover code on this quantitative question was given when students used a conceptual approach for at least one part.

Solutions were coded as *correct* if the correct expression for the relevant force was given, $mg \sin\theta$, and if either the calculation or conceptual approach was correct. Students' correctness score was the sum of their correctness on the two isomorphic calculation questions.

*3. Cued symbolic evaluation question (ballistic pendulum)*

Because the cued symbolic evaluation question explicitly directed students not to perform a calculation, we were stricter in when we gave a conceptual code than we were in the qualitative judgment question (exploding blocks). We coded an approach as *conceptual* only when the student evaluated the given expression against the expected physical behavior. More specifically, an approach was conceptual when (i) the direct and/or inverse proportional dependences in the given expression were tested against the expected physical behavior, or (ii) the speeds before and after collision were compared and tested against the expected physical behavior. The conceptual code is the crossover code here, since the standard version from which this question is adapted is quantitative (i.e., "calculate the symbolic expression").

Solutions were coded as *correct* if the given mathematical expression for angular speed was deemed implausible and a correct approach was taken. A correct use of proportional dependence would reject the given expression because the relations between the rod's mass or the bullet's mass and final angular speed were incorrect. A correct speed comparison would reject this expression because it says that the (linear or angular) speed increased after collision, a physical impossibility. These two conceptual approaches were the only two approaches found to correctly debunk the proposed expression.

*4. Interrater reliability*

The coding scheme was initially generated by three of the authors by examining a small subset of student responses. Then, after initially using another small set of student responses to calibrate their coding, the first and second authors then coded 45% of students' responses to all three crossover problems and their associated standard problems, distributed proportionally across the data collected from the three instructional groups. After this initial round, a second round of coding broke down the approach and accuracy of the crossover problems in greater depth, as reported in the coding scheme. In this second round, the first and second authors recoded a subset of student responses (20% of the total data corpus[2]). After each round of coding, the authors discussed disagreements and modified the coding scheme to resolve those disagreements. For all results presented, the coders reached an average of 95% agreement (average κ = .85; lowest code agreement = 88%, lowest code κ = .75). The first author then coded all remaining responses.

---

[2] For the calculation attempt codes on the isomorphic calculation problems, only 30 responses were coded, because this was determined to be a simple code to apply. Agreement on this code was in-line with the average agreement for all other codes.

## VI. RESULTS

The subsequent analysis excludes the 23 students (5 CTRL students, 15 MS-nov students, and 3 MS students) who did not attempt all three crossover problems, leaving 324 students (CTRL $n = 72$, MS $n = 134$, MS-nov $n = 118$). The exclusion of these 23 students did not change the overall patterns of significance in the results.

### A. Mathematical sensemaking supports calculation-concept crossover approach use

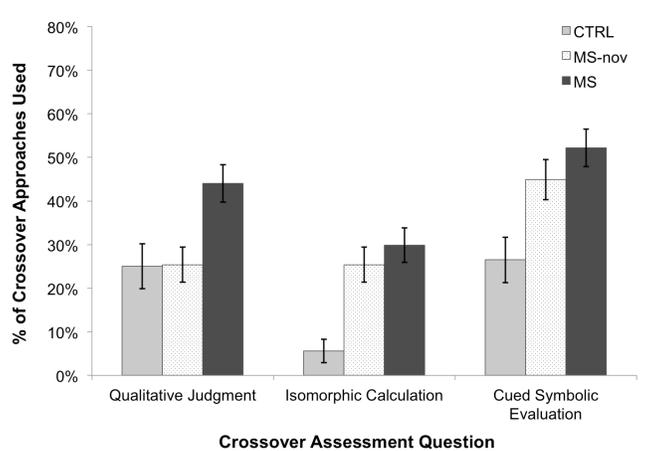

Figure 5. Percentage of calculation-concept crossover approaches used in the three instructional groups on the crossover assessments. Error bars represent ±1 SEM, calculated from the binomial distribution.

Figure 5 shows the percentage of calculation-concept crossover approaches used by the three instructional groups. Our primary comparison of interest is between the MS and CTRL groups. Overall, the MS group used crossover approaches more often than the CTRL instructional group. Compared to CTRL students, MS students spontaneously used more calculations on the qualitative judgment problem, $\chi^2(1, N = 206) = 7.25$, $p = .007$, and more conceptual reasoning on the isomorphic calculation problem, $\chi^2(1, N = 206) = 16.5$, $p < .001$. Even though the cued symbolic evaluation problem told students not to appeal to calculations, MS students still used conceptual reasoning more often than their CTRL counterparts, $\chi^2(1, N = 206) = 12.8$, $p < .001$. This confirmed our main prediction for all three crossover assessments: MS instruction better supported calculation-concept crossover when solving physics problems compared to CTRL instruction.

The MS-nov group partially matched our predictions for the mathematical sensemaking curriculum. While MS-nov students used more crossover approaches than CTRL students for the isomorphic calculation question, $\chi^2(1, N = 190) = 12.0$, $p < .001$, and the cued symbolic evaluation question, $\chi^2(1, N = 190) = 6.52$, $p = .01$, this was not true of the qualitative judgment question, $\chi^2(1, N = 190) < .01$, $p > .90$.

### B. Correctness on crossover assessments and associated standard problems

Turning to the correctness on each of the three crossover assessments (Fig. 6), mathematical sensemaking students generally outperformed CTRL students on qualitative judgment and cued symbolic evaluation. This description fit the MS group exactly, outperforming the CTRL group on the qualitative judgment question, $\chi^2(1, N = 206) = 8.89$, $p = .003$, and the cued symbolic evaluation question, $\chi^2(1, N = 206) = 12.9$, $p < .001$. The MS-nov group outperformed the CTRL group on the symbolic evaluation question, $\chi^2(1, N = 190) = 7.90$, $p = .005$, but not the qualitative judgment question, $\chi^2(1, N = 190) = 1.18$, $p = .28$. We also made the prediction that crossover approach use would not help students be more accurate on the isomorphic calculation questions, since the calculation is relatively straightforward. On these questions, the CTRL group trended to be more correct than either mathematical sensemaking group, but this difference was not significant when comparing to either the MS group, $t(172.1) = 1.35$, $p = .18$, or the MS-nov group, $t(188) = 1.82$, $p = .07$.

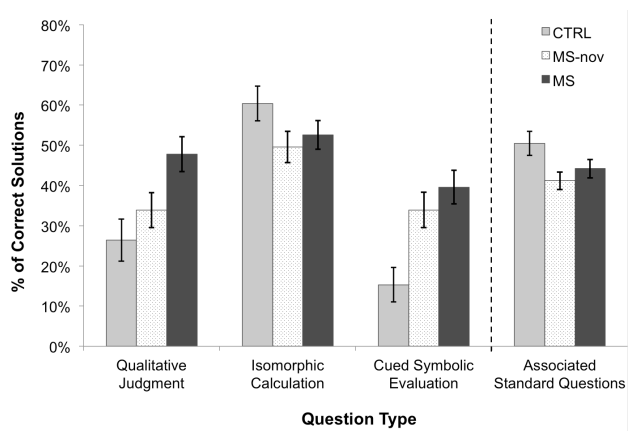

Figure 6. Percentage of correct solutions of the three instructional groups on the three crossover assessments and the associated standard questions. Error bars represent ±1 SEM, calculated from the binomial distribution for binary measures.

One possibility for why the mathematical sensemaking groups outperform the CTRL group on the crossover items is because of generally better physics problem-solving skill in the MS and MS-nov groups. To explore this possible explanation, we compared the three groups on their mean performance on the 6 associated standard problems. There was a difference between the three groups on the total associated standard problem score, $F(2, 321) = 3.03$, $p < .05$. To correct for multiple comparisons, we made pairwise group comparisons with the Games-Howell procedure. The only significant pairwise difference was that the CTRL group scored higher on associated standard problems than the MS-nov students, $p = .04$, $d = .38$. The MS-nov group's worse performance on the standard problems makes their better performance on the cued symbolic evaluation question even more notable. Similarly, even though there was no significant performance difference between the MS and CTRL groups on the standard problems, the MS students were more accurate on the crossover assessments, when predicted. These results indicate that mathematical sensemaking instruction does not benefit students by improving their "general" problem solving skills. Rather, there is a particular benefit that is captured by the calculation-crossover assessments.

Given this conclusion, one follow-up question is whether MS and MS-nov students outperformed CTRL students on the crossover questions *because* they used crossover approaches more frequently on those items. Notably, the patterns of significant differences between groups in correctness in Figure 6 match the patterns of differences for crossover approaches used in Figure 5. On the cued symbolic evaluation question, the connection is an obvious one. The results of the coding revealed that only the two coded crossover approaches, comparing proportional dependencies to physical behavior or comparing initial and final speeds, yielded a correct judgment. Therefore, success on this question is by definition connected to the coded crossover approach use.

However, the connection between crossover approaches and correct answers on the qualitative judgment problem bears closer analysis. In principle, both conceptual reasoning and calculation use (the crossover approach here) can yield the correct answer. Figure 7 breaks out the approach categories coded (calculation only, conceptual reasoning only, both calculation & conceptual reasoning) for the three instructional groups and indicates what percentage of solutions taking each approach was correct. The crossover approach percentage shown in Figure 5 is the sum of "calculation only" and "calculation & conceptual reasoning" approaches. On this problem, only 19% of conceptual reasoning only approaches were correct. In comparison, approaches that included a calculation were much more successful.

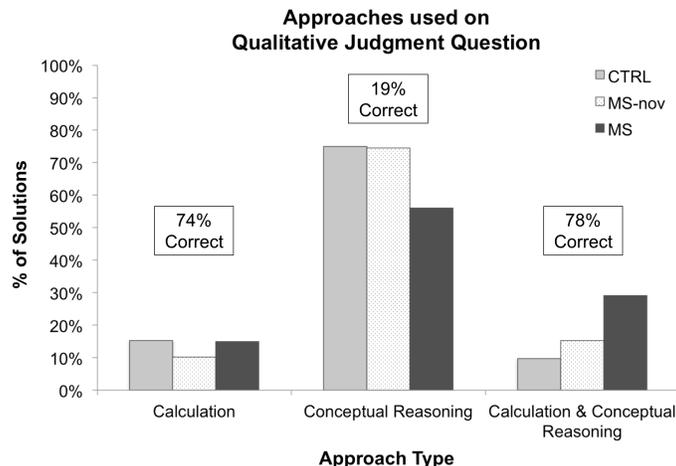

Figure 7. Percentage of each approach type – calculation, conceptual reasoning, and both calculation & conceptual reasoning – used on the qualitative judgment problem along with the percentage of correct answers produced by each approach.

A breakdown of the common conceptual reasoning errors shows that this low success rate comes from misapplications of common explanations in introductory physics. 49% of these conceptual reasoning only approaches concluded that the mechanical energy will remain the same. The two most common justifications were (i) the general principle "energy is always conserved" and (ii) a compensation argument: the energy lost by the stopped block would be exactly gained by the accelerated block, leaving the overall mechanical energy the same. 19% of conceptual reasoning only approaches concluded that energy would decrease, commonly citing non-mechanical energy released by the system in the explosion (e.g. heat, light, sound, deformation, etc.). The remaining errors were incorrect justifications of the correct final answer or solutions that left the final answer ambiguous.

This breakdown suggests that the MS group's increased use of crossover approaches – here, calculations – and the higher correctness rate of approaches that incorporated a calculation explains why the MS group was more correct on the qualitative judgment problem than the CTRL group. To test this mediation, we performed a 3x2x2 log-linear analysis, using instruction (CTRL, MS-nov, or MS), crossover approach use (Did or did not include a calculation), and correctness (correct or incorrect) as the three factors. Log-linear analysis tests for relationships between multiple categorical variables. Our analysis used log-linear model selection, which starts with a completely saturated model, including all 1-way, 2-way, and 3-way relationships and removes the highest-order relationships that do not significantly contribute to the fit of the model, one at a time until all remaining terms contribute significantly to the model. The final model contains only the highest-order, significant relationships between factors. In the first step of model selection, the 3-way association was deemed to not significantly contribute to the fit of the model, $\chi^2_{change}(2, N = 324) = 1.63$, $p = .44$, and was removed. This indicated that the percentage of correct answers produced by each approach did not differ by instructional group. In the second step, the relationship between instruction and correctness was removed, $\chi^2_{change}(2, N = 324) = 3.26$, $p = .20$. This indicated that there was no direct association between instructional group (CTRL vs. MS vs. MS-nov) and correctness. In the final model, instruction was associated with approach, $\chi^2_{change}(2, N = 324) = 12.4$, $p = .002$, and approach was associated with correctness, $\chi^2_{change}(1, N = 324) = 104$, $p < .001$, confirming that the link between MS group and correct answers is mediated by crossover approach use. The final overall model fit did not significantly deviate from the data, $\chi^2(4, N = 324) = 4.90$, $p = .30$.

### C. Mathematical sensemaking supports explicit coherence-seeking

The mathematical sensemaking instruction fosters coherence-seeking between calculations and conceptual reasoning. Crossover approaches are hypothesized to be supported by that coherence-seeking, at least tacitly. Alternatively, we can look for explicit demonstrations of coherence-seeking through solutions giving both a calculation and a conceptual reason for an answer. Table 4 shows the percentage of approaches that demonstrated this kind of explicit coherence-seeking, omitting the cued symbolic evaluation question because it prompted students not to use a calculation. The explicit coherence-seeking approaches on the qualitative judgment question are just a renaming of the "calculation & conceptual reasoning" approaches shown in Figure 7.

Table 4. The percentage of explicit coherence-seeking approaches (both calculation and conceptual reasoning) used on the qualitative judgment question and the isomorphic calculation questions. * indicates percentage is greater than CTRL percentage, $p < .01$.

| | **Explicit Coherence Approaches** | |
|---|---|---|
| | Qualitative Judgment Question | Isomorphic Calculation Questions |
| CTRL | 9.7% | 2.8% |
| MS-nov | 15.3% | 16.1% * |
| MS | 29.1% * | 14.9% * |

In sum, the patterns of explicit coherence approaches mirror the patterns for crossover approach use. MS students gave more explicit coherence-seeking responses than CTRL students on the qualitative judgment question, $\chi^2(1, N = 206) = 10.1$, $p = .001$, and the isomorphic calculation question, $\chi^2(1, N = 206) = 7.25$, $p = .007$. The MS-nov group did not display explicit coherence-seeking more than the CTRL group on the qualitative judgment question, $\chi^2(1, N = 190) = 1.20$, $p = .27$, but they did on the isomorphic calculation questions, $\chi^2(1, N = 190) = 8.07$, $p = .004$. This illustrates the success of the mathematical sensemaking instruction for having students explicitly demonstrate coherence between calculations and concepts in their solutions.

Although standard problem-solving frameworks often include a "check your answer" step at the end, they differ from our focus on explicit coherence by making calculations primary and other approaches a secondary check of that calculation. Our focus on coherence-seeking between calculations and conceptual reasoning places the emphasis on the coherence rather than the primacy of one approach over another. This is more descriptive of a wider range of problem-solving approaches, as an initial conceptual explanation may *precede* the calculation rather than follow it.

### D. Associations with problem-solving epistemologies: MPEX2 results

On top of completing the crossover assessments, 240 students (CTRL: $n = 47$; MS-nov: $n = 94$; MS: $n = 99$) also completed the modified version of the MPEX2 (leaving no more than 2 out of 32 items blank) in the final week of the course. Before analyzing the results, we selected 15 MPEX items that were tied to the mathematical sensemaking instructional goals of fostering coherence-seeking and problem-solving flexibility. Favorable responses to these items had a high reliability ($\alpha = .82$) and were combined into an MS epistemology score (percentage of favorable responses, ranging from 0 to 100%). The averages for each instructional group are shown in Figure 8.

There was a significant difference between MS epistemology scores by instruction, $F(2, 237) = 19.6$, $p < .001$. Post-hoc comparisons using the Games-Howell test reveal that MS-nov students scored higher than CTRL students, $p < .001$, $d = .70$, and MS students scored higher than MS-nov students, $p = .007$, $d = .44$ (Implying, of course, that MS students scored higher than CTRL students, $p < .001$, $d = 1.18$). Notably, no CTRL students had an MS Epistemology score above 60% whereas 22% of MS-nov students and 43% of MS students did.

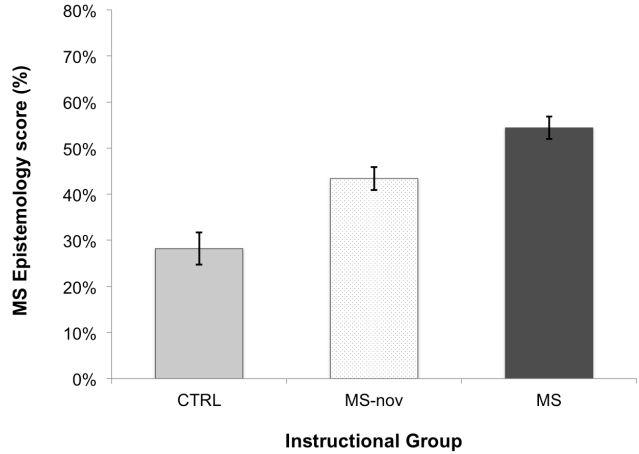

Figure 8. Average MS Epistemology score (% of favorable responses) for each instructional group. Error bars represent ±1 SEM.

These results suggest one possibility: the increased use of crossover approaches by the MS and MS-nov students is explained by their MS epistemology score. That is, the mathematical sensemaking instruction impacts students' problem solving by changing their espoused views on how calculations and concepts should be used together in problem solving. To test the relation between MS epistemology score and the number of crossover approaches used, we summed the number of crossover approaches used by each student, ranging from 0 – 3. Because the distribution of crossover approaches used was skewed toward zero, we used a Poisson-distributed general linear model with the number of crossover approaches as the dependent variable and instructional group and MS epistemology score as the independent variables. Each student's crossover approach total is modeled through the equation

$$y = exp[C + b_{MS-nov}x_{MS-nov} + b_{MS}x_{MS} + b_{MS-epistemology}x_{MS-epistemology}]$$

where $y$ is the number of crossover approaches a student used, $x_{MS-nov}$ is 1 for students in the MS-nov group and 0 otherwise, $x_{MS}$ is 1 for students in the MS group and 0 otherwise, and $x_{MS-epistemology}$ is a student's MS-epistemology score (ranging in percentage from 0 to 100), and the $b$'s are the associated model coefficients for the $x$'s.

The model is plotted in Figure 9 and the model coefficients are shown in Table 5. The model fit for the CTRL group is only shown for MS epistemology scores from 0% of 60%, because no CTRL students score outside of this range. The coefficient for MS epistemology score is significant and positive. This supported the prediction that epistemological views favoring coherence between calculations and concepts are associated with crossover approach use. In addition, compared to CTRL students, MS students used significantly more crossover approaches even after controlling for MS epistemology score. The difference between MS-nov and CTRL groups was not significant. This indicated that even for CTRL and MS students who received the same MS epistemology score, MS students used more crossover approaches. Said another way, MS epistemology score alone does not explain the increased use of crossover approaches by MS students. The pattern of significant results also holds when excluding all students with MS epistemology scores greater than 60%, testing only the region of overlap between all three instructional groups. A model which included an interaction between MS epistemology score and instruction group was tested, and the interaction was found to be non-significant.

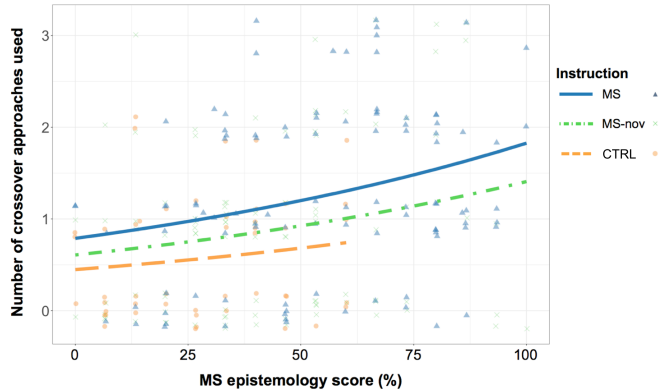

Figure 9. Scatterplot of number of crossover approaches used vs. MS epistemology score, with the model fit plotted for each instructional group.

Table 5. The coefficients of the general linear model with number of crossover approaches used as the dependent variable and instructional group and MS epistemology score as the independent variables.

| Factor | B | se | Z | p |
| --- | --- | --- | --- | --- |
| Constant | -0.80 | 0.21 | 3.83 | < .001 |
| Instructional group | | | | |
|   CTRL | ----- | ----- | ----- | ----- |
|   MS-nov | 0.31 | 0.23 | 1.35 | 0.18 |
|   MS | 0.57 | 0.23 | 2.51 | 0.01 |
| MS epistemology score | 0.0084 | 0.0027 | 3.11 | 0.002 |

## VII. DISCUSSION

On one hand, this study illustrates the benefits of the mathematical sensemaking curriculum, focused on developing coherence seeking between calculations and conceptual reasoning and epistemological views that buttress such coherence-seeking. This two-pronged focus on coherence changed how students approach problem solving, increasing the use of calculation-concept crossover approaches. Overall, these crossover approaches were more efficient, effective, and insightful than the common alternatives. Mathematical sensemaking instruction also increased explicit demonstrations of coherence, providing additional evidence for the coherence-seeking outcomes of this instructional approach. Epistemologically, the mathematical sensemaking instruction was more successful at leading students to espouse epistemologies favoring coherence-seeking, and these epistemologies were also associated with the more flexible calculation-concept crossover approaches. Compared to the CTRL group, the predicted effects all bore out exactly for the MS group and partially for the MS-nov group, showing that even an instructor teaching a large lecture course for the first time can achieve some of the positive benefits of the mathematical sensemaking curriculum.

On the other hand, this study also introduces calculation-concept crossover as an assessment framework that highlights some of the benefits of mathematical sensemaking. If we aim to teach students the mathematical sensemaking skills necessary for future problem solving, searching for calculation-concept crossover can be a valuable assessment target. Even as the taxonomies of physics problem-solving approaches grow [21,50–53], this is the first study we know of that systematically measures how a novel instructional method develops beneficial problem-solving flexibility in physics. Instruction based on the standard problem-solving paradigm has successfully structured students' problem solving around initial conceptual analysis before diving into mathematical manipulations but has not explicitly attended to the coherence-seeking benefits seen in calculation-concept crossover. Similarly, most standard problem-solving rubrics in physics do not attend to the difference between calculation and conceptual approaches on quantitative problems [34].

## A. Testing calculation-concept crossover and standard problem solving: an example of multidimensional assessment

Both the experienced and novice mathematical sensemaking instructors' students used more crossover approaches, gave more correct answers on crossover questions, and espoused stronger coherence-favoring epistemologies than the CTRL students. However, the MS-nov students performed significantly worse on standard quantitative problems than the CTRL students, whereas the MS students did not. One obvious question is: for the novice instructor, did the gains in mathematical sensemaking come at the expense of standard problem-solving skills?

This question cannot be fully addressed empirically in this study. One reason is the lack of a CTRL-nov group in our study, which would allow for an estimate of how lower instructor experience impacts standard problem-solving skills independent of the instructional approach. Yet, even with such a group, there are many uncontrolled differences across the courses in this study that are not captured by the labels CTRL, MS-nov, and MS. Teaching is complex [54] and differences between the courses on a variety of instructional dimensions – such as classroom management strategies, relationships between students and instructors, or how the instructors' teaching styles embody their values and beliefs about teaching – could all be playing a role in how these results emerge. More classroom research is needed to see if the patterns in learning outcomes seen here rise above the noise of these instructional variations.

Although this study does not settle the question of whether the MS-nov students represent an instructional trade-off between different learning outcomes, note that assessments investigating multiple dimensions of learning are required to ask this question in the first place. Methodologically, we argue that this study embodies an approach to multidimensional assessment that is necessary for evaluating the success of instructional approaches on multiple learning goals. Even when an instructional approach shows significant gains for one learning goal, it is important to know its effect on other learning goals, whether it is positive, negative, or neutral.

Historically, PER has usually emphasized the ways in which new instructional methods can improve learning for one learning goal, such as conceptual understanding, without compromising other goals, such as quantitative problem-solving skill. For this reason, the idea of trade-offs between multiple learning goals has not been addressed. For example, in the case of active learning versus traditional lecture, meta-analytic studies have shown that active learning leads to greater learning gains along multiple course objectives, including better exam scores, better scores on concept inventories, and lower class failure rates [55]. In this case, the data suggest that active learning approaches are strictly better than traditional lecture on a variety of multiple educational goals. However, little is known about the comparative benefits of different active learning environments, and as PER investigates these finer-grained instructional differences, the possibility exists that promoting learning in one direction leads to trade-offs in another. At some point, instructional decisions may need to rest on decisions of value: what outcomes do I value more and what outcomes do I value less. In the case of the two types of assessments in our study, some instructors may be willing to risk sacrificing some levels of basic problem-solving competence for increased mathematical sensemaking; others may not. Research assessing multiple dimensions of learning can illuminate these potential debates.

Additionally, having multiple types of assessments can allow for testing of as-of-yet untested empirical questions that could inform those value judgments. For example, how do the two dimensions of problem solving investigated in this study support learning and success in future STEM courses and careers? How do these problem-solving skills (along with others) at the introductory level seed students' trajectories toward expertise? Another key question is about the effects of these different learning outcomes on retention and persistence – in physics specifically and STEM fields more generally. Students wanting to engage in key disciplinary sensemaking practices, such as coherence-seeking and mathematical sensemaking, may lose interest in STEM domains if they are primarily training on more routine problem-solving competencies [56]. In this way, early exposure to mathematical sensemaking may prove to be more valuable in the long run to students' educational trajectories. Longitudinal hypotheses like these often go untested, because they require methodological power, serious time investment, and, as we argue, multiple types of assessment.

Although any one assessment highlights a dimension of learning, relying too heavily on any single assessment may obscure potential learning or missed learning opportunities along other dimensions. In addition to assessing the potential learning benefits of an instructional approach, researchers should also seek to accurately assess potential trade-offs. This will demand a nuanced look at instructional comparisons and require multidimensional assessments. The potential payoff will be better empirical data for making informed decisions about instructional goals and better methodological tools for doing research that can inform these decisions.

### B. The role of students' physics classroom expectations

Part of the instructional scaffolding in the mathematical sensemaking curriculum is that students learn crossover reasoning is rewarded. Throughout the semester, the mathematical sensemaking courses valued student use of multiple problem-solving approaches and seeking coherence between them. The grading on midterm exams reflected this, as students could receive some credit for articulating multiple approaches, seeking coherence (or noting unresolved incoherence) between these approaches, and articulating intuitive insights, even if the final answer or approach was incorrect. It is plausible that this scaffolding of students' expectations contributed to the difference in students' espoused views on problem solving, as measured by the MPEX2.

As with all such instructional scaffolding, the key question is whether the changes in learning and performance will persist when that explicit scaffolding is not present. If future environments do not explicitly encourage or reward the aspects of problem-solving flexibility, will students persist in using them? The degree to which the mathematical sensemaking instruction impacts students' thinking and learning in future educational and professional contexts is an important direction for future research.

A more immediate question is how this "expectation scaffolding" should inform the interpretation of this study's results. On the final exam, which contained the crossover and standard problem-solving questions, we find it plausible that students in the mathematical sensemaking courses, on average, had a greater expectation that crossover approaches, and seeking coherence between calculations and conceptual reasoning more generally, would be rewarded on the final exam. Therefore, we should interpret the crossover assessment results as indicating how students approach problems in the environment of the mathematical sensemaking course, not what students' take away from the course. Again, in our view the difference in grading expectations is a key part of the instructional scaffolding of the mathematical sensemaking instruction that helps students practice and enact coherence seeking and mathematical sensemaking processes.

That said, although students' expectations of how to get a good grade shape their behavior in the course, these expectations alone cannot explain the difference in reasoning we see between the different classes. In any physics class, students expect that answering questions on an exam correctly will lead to the best grade, but students in the mathematical sensemaking courses were more correct overall on two of the three crossover assessments (as predicted). One of these crossover assessments, the cued symbolic evaluation question (ballistic pendulum), even explicitly instructed students *not* to rely on calculation, implying the productivity of a crossover approach. Therefore, we argue that in addition to mathematical sensemaking students' greater expectations that coherence-seeking approaches will be valued, the mathematical sensemaking curriculum also developed their skills for employing that coherence for producing correct solutions.

The relationship between surveyed epistemology and number of crossover approaches used also shows how current measures of epistemologies/expectations fail to explain the difference in crossover approaches between the CTRL and MS students. Controlling for their epistemological scores, the MS students used more crossover approaches than the CTRL students. This indicates that comparing students with the same epistemology score, MS students used more crossover approaches than CTRL students. We can hypothesize at least three factors that can explain the remaining difference between the MS and the CTRL instruction. One is the previously stated explanation, that this difference represents differences in the coherence-seeking skills developed in the MS and CTRL instruction. A second possibility is that the MS epistemology score is too broad and does not accurately capture differences in students' views relevant for crossover approach use. In this case, there would be unmeasured dimensions of students' epistemological views or expectations that could more completely explain the difference in crossover approach use between MS and CTRL groups. A third possibility is that students' surveyed epistemologies may not accurately reflect the epistemologies-in-use developed in the MS instruction that more directly drive problem solving. We expect that all three of these factors contribute to the remaining difference in crossover approach use between MS and CTRL groups to some degree, but how much each of these explanations (or others) contribute to this difference is a question for future research.

### C. The mistake in labelling qualitative questions as "conceptual questions" and quantitative questions as "calculation questions"

This study expands the discussion around how typical types of quantitative and qualitative physics assessment questions should be interpreted and designed. In PER, it is still standard (and productive) to treat quantitative questions as assessments of calculation skill and qualitative questions as assessments of conceptual knowledge. This is embodied with quantitative questions by the fact that quantitative problem-solving rubrics are designed to assess the quality of the calculation used and are not well suited to assess purely conceptual approaches. On the other side, banks of qualitative questions are often explicitly labeled as conceptual questions, such as in the Force *Concept* Inventory or the *Conceptual* Survey of Electricity and Magnetism. In many cases, these classifications are wholly accurate. On many typical quantitative questions requiring a precise numerical or symbolic result, calculations are necessary (e.g., finding the final velocity of a block sliding down a ramp while experiencing friction, given the relevant ramp parameters). Similarly, many qualitative questions require conceptual understanding and may not be indicated through calculations (e.g., naming the forces acting on a block sliding down a ramp while experiencing friction).

Yet, we have shown that these standard interpretations would not accurately capture students' reasoning on our crossover assessments. On the qualitative judgment problem, the *calculation* approaches were more likely to yield correct answers. In these cases, it would an error to interpret the correct answers as indicating *conceptual* knowledge. Similarly, on the isomorphic calculation problem, it would be an error to interpret correct answers as only indicating calculation skill, since many students found a conceptual similarity between problems that helped them determine the answer. These interpretations are consequential, because they lead to different instructional implications. For example, interpreting poor performance on qualitative questions as weak conceptual understanding suggests that conceptually-focused instruction is needed. However, in the framework of calculation-concept crossover, performance can be improved—at least in some cases—by helping students see the usefulness of calculations for qualitative problems.

To further the investigation into calculation-concept crossover, new assessment items will need to be designed. Simply searching for crossover approaches on standard problems that do not afford crossover solutions will not be fruitful. Here, in contrast with the associated standard problems, we used three assessment questions carefully designed to illicit and detect crossover approaches. Although we expect that some standard problems will invite students to seek coherence between different problems and different components of their knowledge, many will not, so theoretical elaboration of this space will require the careful design of new assessment questions.

### D. Calculation-concept crossover demonstrates adaptive expertise

This work takes a step towards understanding adaptive expertise in physics education. Hatano and Inagaki [57] distinguished routine and adaptive expertise: while routine expertise involves using standard approaches in familiar situations, adaptive expertise allows people to find new solutions to new problems. This adaptation can involve modification of known procedures or invention of novel approaches. We contend that calculation-concept crossover marks adaptive expertise. On our novel crossover assessments, students could break from standard approaches (i.e., calculations on quantitative problems or conceptual reasoning on qualitative problems) to find more efficient, effective, and elegant solutions. Although our crossover assessments are not so far from the typical problem space of introductory physics, we believe that students using crossover approaches here demonstrate adaptability and flexibility that could forecast success in adapting to new problems in the future.

Schwartz, Bransford, and Sears [58] broke adaptive expertise into two components: efficiency and innovation (Fig. 10). Importantly, while both routine and adaptive expertise can behave efficiently in familiar settings, it is innovation that differentiates these two courses of expertise. Considering these two dimensions together, they argue that both efficiency and innovation should proceed together in the development of adaptive expertise, giving examples of how a focus on just one or the other is not as successful at helping students transfer their knowledge to new situations. They hypothesize an *optimal adaptability corridor* that balances both efficiency and innovation in instruction.

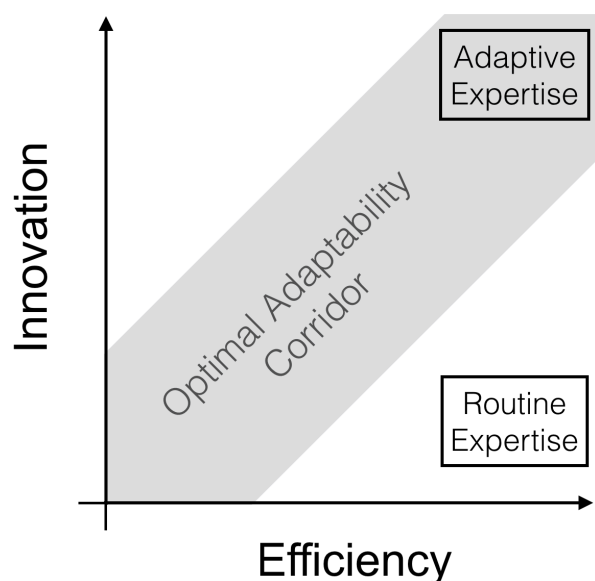

Figure 10. Two courses of expertise plotted on a 2-D space of innovation vs. efficiency (adapted from Schwartz, Bransford, and Sears [58].

We can fit this efficiency-and-innovation framework to interpret our study in two ways. In terms of instruction, we propose that the promotion of student sensemaking in the lecture provided important innovation experiences in mathematical sensemaking classrooms. Rather than directing students to the correct reasoning as efficiently as possible, the episode in section IV.B illustrates how the mathematical sensemaking instruction gave students an opportunity to invent explanations from their own reasoning. These opportunities to invent and be innovative are a key part of the mathematical sensemaking curriculum and aim to foster the skills, experience, and dispositions students will need to be innovative in future settings. While standard, lecture-based instructional approaches seem to focus on efficiency, many active-engagement instructional approaches balance content learning with opportunities for student innovation and invention. This suggests that many existing PER-based instructional methods may offer opportunities to develop adaptive expertise, and it would be an interesting direction for future research to investigate how these existing instructional approaches might help foster that expertise.

In terms of the learning outcomes, we can map standard problem accuracy as being a measure of routine efficiency and crossover approaches as being a measure of innovation (i.e. identifying and choosing to use a productive crossover approach rather than a more standard approach). Figure 11 plots this mapping for the three instructional groups in our study, where the dotted line represents equal success on efficiency and innovation measures. Although this mapping shouldn't be taken too seriously since the exact percentages depend as much on the comparative difficulty of the standard and crossover assessment problems as they do on the status of students' problem-solving skill, this cartoon provides a starting point for thinking about educational possibilities in introductory physics.

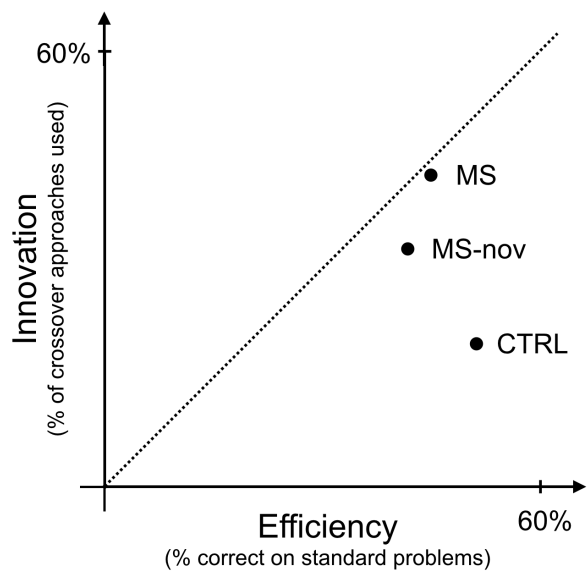

Figure 11. Plot of the study results by innovation vs. efficiency. The dotted line represents a balance of equal levels of efficiency and innovation. The MS student average is closest to this line whereas the CTRL student average is furthest from this line.

Specifically, these results provide a counter to the educational idea that training for adaptive expertise can only occur after a sufficient amount of routine expertise is developed. This trajectory is embodied by introductory teaching that focuses on basic skill development, testing for their efficient use on familiar problem types, and saving adaptivity and innovation for future courses that can leverage this earlier instruction. In this case, the introductory courses teach the basic skills, and the upper-division courses will later provide opportunities to evolve those basic skills into the adaptive skills that constitute "thinking like a physicist." Results from the control classroom in our study illustrate the expected outcomes of this "efficiency-before-innovation" model. By contrast, the results of the mathematical sensemaking instruction show that, even in introductory courses, aiming to balance efficiency and innovation can be fruitful. The MS group showed higher levels of crossover approach use than the CTRL group while demonstrating a comparable level of performance on the standard problems. Here, the mathematical sensemaking instruction illustrates the possibility of effectively developing "efficiency-alongside-innovation."

It might also be hypothesized that the ability to develop efficiency-alongside-innovation relies on instructor experience, but the performance of the MS-nov students counters this hypothesis. Rather, the MS-nov group shows that even first-time large-lecture instructors may be able to balance efficiency with innovation to some degree, indicated by the shorter relative distance to the dotted line in Figure 11 representing equal levels of efficiency and innovation, even though the MS-nov group performed worse than the CTRL group on standard problems representing routine efficiency.

When aiming to develop adaptive expertise, training for efficiency-before-innovation may be taking the long way around. On the one hand, doing so misses opportunities to foster innovation early on, so that skills for invention can develop along with skills for efficiency. On the other hand, training only for efficiency may result in over-routinization. The overzealous use of routines in new situations may overshadow opportunities for innovation and novel exploration [59], as they did for students who did not leverage the isomorphism between problems on the isomorphic calculation questions and for students who did not employ calculations on the qualitative judgment question. Attempts to help students learn the standard procedures may even cue students to avoid non-standard methods and, inadvertently, suppress the search for new, more efficient and elegant approaches [60], raising potential barriers to future innovation.

## VIII. CONCLUSION

The goal when teaching problem solving is to provide students with the skills they need to adapt to the new, unforeseeable problems they will face in their educational and professional futures. Solving these future problems will require that students are able to draw on their procedural knowledge and skills for efficiency as well capitalize on opportunities to adapt to new situations with new interpretations, methods, and approaches. We propose that this adaptation in physics is supported by coherence: coherence between calculations and physical concepts as well as coherence between how one approaches quantitative and qualitative problem solving. Teaching introductory physics with a focus on mathematical sensemaking through coherence seeking may help us discover new efficiencies in the teaching of physics, though evaluating its success will require innovations in assessment.

**Acknowledgements**

The research described here was supported by NSF EEC-0835880.